\documentclass[twocolumn,showpacs,prl,superscriptaddress]{revtex4-1}%
\usepackage{amsfonts}
\usepackage{amsmath}
\usepackage{amssymb}
\usepackage{mathrsfs}
\usepackage{graphicx}
\usepackage{txfonts}
\usepackage{xcolor}%
\usepackage{ulem}
\providecommand{\U}[1]{\protect\rule{.1in}{.1in}}

\begin{document}
\title{Detection of quantum \textcolor{black}{signals} free of classical noise via quantum correlation}

\author{Yang Shen$^{\S}$}
\affiliation{Department of Physics and the IAS Center for Quantum Technologies, The Hong Kong University of Science and Technology, Clear Water Bay, Kowloon, Hong Kong, China}

\author{Ping Wang$^{\S}$}
\email{wpking@bnu.edu.cn}
\affiliation{College of Education for the future, Beijing Normal University, Zhuhai 519087, China}
\affiliation{Department of Physics, The Chinese University of Hong Kong, Shatin, New Territories, Hong Kong, China}
\affiliation{Centre for Quantum Coherence and The Hong Kong Institute of Quantum Information Science and Technology,
The Chinese University of Hong Kong, Shatin, New Territories, Hong Kong, China}

\author{Chun Tung Cheung}
\affiliation{Department of Physics, The Chinese University of Hong Kong, Shatin, New Territories, Hong Kong, China}

\author{J\"{o}rg Wrachtrup}
\affiliation{3. Physikalisches Institut, Research Center SCOPE and Integrated Quantum Science and Technology (IQST), University of Stuttgart, Pfaffenwaldring 57, 70569 Stuttgart, Germany}

\author{Ren-Bao Liu}
\email{rbliu@cuhk.edu.hk}
\affiliation{Department of Physics, The Chinese University of Hong Kong, Shatin, New Territories, Hong Kong, China}
\affiliation{Centre for Quantum Coherence and The Hong Kong Institute of Quantum Information Science and Technology,
The Chinese University of Hong Kong, Shatin, New Territories, Hong Kong, China}

\author{Sen Yang}
\email{phsyang@ust.hk\\
$^{\S}$ These authors contributed equally to this work}
\affiliation{Department of Physics and the IAS Center for Quantum Technologies, The Hong Kong University of Science and Technology, Clear Water Bay, Kowloon, Hong Kong, China}
\affiliation{Department of Physics, The Chinese University of Hong Kong, Shatin, New Territories, Hong Kong, China}

\begin{abstract}
Extracting useful signals is key to both classical and quantum technologies.
Conventional noise filtering methods rely on different patterns of
signal and noise in frequency or time domains, thus \textcolor{black}{limiting} their scope
of application, especially in quantum sensing. Here, we propose a
signal-nature-based (not signal-pattern-based) approach which singles
out \textcolor{black}{a quantum signal} from its classical noise background by employing
\textcolor{black}{the intrinsic quantum nature of the system}. We design a novel protocol
to extract the \textit{Quantum Correlation} signal and use it to single
out the signal of a remote nuclear spin from its overwhelming classical
noise backgrounds, which is impossible to be accomplished by conventional
filter methods. Our work demonstrates the quantum/classical nature
as a new degree of freedom in quantum sensing. The further generalization
of this quantum nature-based method opens a new direction in quantum
research.
\end{abstract}

\maketitle


Digital signal processing (DSP) and noise filtering techniques are
the foundation of classical information technology\cite{Saeedbook2008}.
These methods, such as transform-based signal processing, model-based
signal processing, Bayesian statistical signal processing and neural
networks, design filters based on the specific pattern of noise in either
spectrum or time domain. These methods also play a crucial role in
the development of quantum information science, as the separation
of quantum signals from strong classical noise has wide applications
ranging from quantum sensing\cite{ZhaoNatNano2012,BylanderNatPhys2011,DegenRMP2017,MenesesPRAP2022},
quantum biology\cite{QinNM2016}, quantum many-body physics\cite{SchweiglerNature2017,ChoiNature2017}
and quantum computing\cite{Nielsenbook2012}. However, these pattern-based
noise filtering methods fail under various circumstances, such as
pattern-less noise (white noise), or when noise is overwhelming in
time or frequency domains, or strong non-stationary noise background\cite{WangCPL2021}.
These circumstances are common in quantum systems, as the interaction
between the sensor and the quantum target is usually weak and buried by classical
noises. Here, we solve this challenge \textcolor{black}{with} a pattern-independent and
noise-free sensing of the quantum target by employing its intrinsic non-commuting
quantum nature despite strong classical noises concealing the signal
of the quantum target. 

\begin{figure}
\includegraphics[width=\columnwidth]{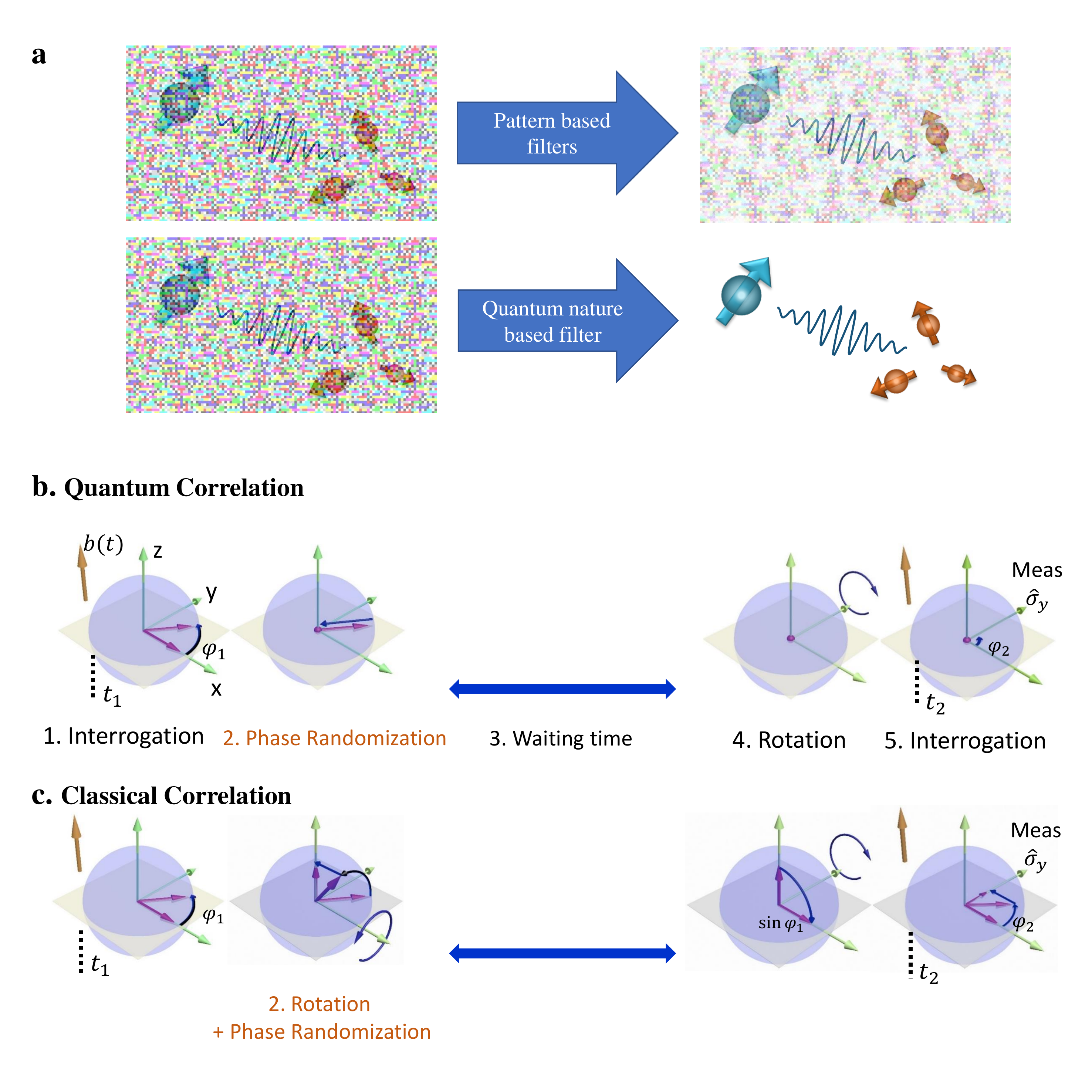}
\caption{$\textbf{Illustration of the quantum nature-based method.}$ Comparison
between traditional DSP methods which are based on pattern-dependent
noise filters (Upper), and the quantum-nature-based filtering method
(Lower). The latter can filter out arbitrary classical noise without
resorting to a specific pattern of noise in the frequency/time domain. $\textbf{b.}$
The semi-classical picture of the \textit{QC} protocol and $\textbf{c}$.
The semi-classical picture of the \textit{CC} protocol. The difference
between the \textit{QC} and \textit{CC} protocols is that the rotation
operation is absent in the \textit{QC} protocols. The final signal
of these protocols is the expectation value of $\hat{\sigma}_{y}$
of the sensor after these operations. Phase randomization denotes the incoherent
operation that eliminates the x-y components of the sensor while keeping
its $z$ components.  \label{Figure1}}
\end{figure}

The signal of the classical entity can be described by a time-dependent
stochastic field $B(t)$, which are always commuted to each other, namely, $[B(t),B(t^{\prime})]=0$. Therefore, Conventional DSP methods
can only differentiate the signal from the noise \textcolor{black}{by} employing different
patterns between the signal and noise (upper graph of Fig. \ref{Figure1}
a). In contrast to the classical entity, the signal of the quantum entity
is originated from a quantum operator $\hat{B}(t)$ which acts on the quantum sensor\cite{WangCPL2021,ZurekPRD1981,BrunePRL1996,MyattNature2000,CywiPRB2008,KorotkovPRB1999,PfenderNC2019}.
The non-commuting nature of the quantum operator $\hat{B}(t)$ can
generate a \textit{quantum signal}, which is absent for the classical
entity. Thus, one can use this non-commuting quantum nature of $\hat{B}(t)$
to single out the signal of the quantum target from any classical
noise background, which always commutes to each other, without the
requirement of any knowledge of the classical noise (down graph of
Fig. \ref{Figure1} a). It should be emphasized that the quantumness
discussed here is defined as the quantumness \textcolor{black}{originating} from the non-commuting
nature of the quantum operator $\hat{B}(t)$, which should be distinguished
from the other quantumness defined in various literature. 

Recently, the non-commuting nature of the quantum target is quantified
by a type of \textit{quantum signal} derived from the time commutator
of a quantum operator, which is called \textit{Quantum Correlation (QC)
}\cite{WangPRL2019,WangCPL2021}. It was proposed that the time correlations
of weak measurements could systematically extract these \textit{QCs}\cite{WangPRL2019}
and filter out arbitrary classical noise backgrounds without resorting
to their specific pattern \cite{WangCPL2021}. However, the experimental
realization of this weak-measurement-based approach is challenging
due to its requirement \textcolor{black}{for} well-readout properties of the quantum sensor.

In this article, we propose a novel protocol to extract the \textit{QC}
and use it to demonstrate, for the first time, to remove the classical
noise background without resorting to the specific pattern of noise.
In contrast to the weak-measurement-based method\cite{WangCPL2021},
this approach does not require multi-times of weak measurements and
hence is easier to realize experimentally. By employing both coherent
and incoherent operations of the sensor and only a one-time readout
of the qubit, we extract an intrinsic\textsl{ }\textit{quantum signal}
from a quantum target, which always vanishes in the semi-classical environment.
Furthermore, we use this protocol to filter out the classical noise
background and realize the \textcolor{black}{pattern-independent} classical noise-free
sensing.

The \textit{QC }signal is extracted by the \textit{QC} protocol as
shown in Fig. \ref{Figure1} b. To analyze the \textit{QC} signal, one
needs to theoretically model the interaction between the quantum sensor
and the target. The semi-classical theory\cite{AndersonRMP1953,AndersonJPSJ1954,WitzelPRB2014,MaPRB2015,YangRPP2017}
models the effects of the environment approximately via a time-dependent
stochastic field. Instead, the quantum theory\cite{ZurekPRD1981,BrunePRL1996,MyattNature2000,CywiPRB2008,WangNC2021}
uses quantum operators to describe the interaction between the quantum
sensor and the quantum target. Below we use both methods to analyze the
\textit{QC} signal.

To illustrate how the \textit{QC} protocol method works, let us introduce
an intuitive picture based on semi-classical theory. As shown in Fig.
\ref{Figure1} b, a quantum sensor, \textcolor{black}{modeled} as a two-level system,
is initialized to $\vert x\rangle$ and then interacts with the environment
for a short time $t_{\mathrm{I}}$ (interrogation process denoted
by Step. 1 in Fig. \ref{Figure1} b). In the semi-classical model,
this environment is treated as a classical field $b(t)$ (Fig. \ref{Figure1}
b). After this step, the sensor acquires a phase $\varphi_{1}\approx b(t_{1})t_{\mathrm{I}}$
and hence the information encoded in phase $\varphi_{1}$ is stored
in the coherence of the sensor. \textcolor{black}{Then a phase randomizing step (Step. 2 in Fig. \ref{Figure1} b) is introduced to eliminate the information stored in the phase of the sensor spin. A random phase is generated on the sensor spin on top of the original phase $\varphi_{1}\approx b(t_{1})t_{\mathrm{I}}$ while keeping its population after this step. Therefore, all possible information about the environment encoded to the sensor in the frame of semi-classical theory is removed. As a result, the information about the environment can never be extracted no matter how the sensor is operated after this step as long as the semi-classical theory holds (e.g. if the quantum sensor is coupled with a classical field nature noise). However, in quantum theory, the entanglement between the quantum sensor and environment introduces back-action to the environment and this back-action information is also encoded in the sensor's population via this entanglement. Consequently, the information about the environment can still be extracted even when phase $\varphi_{1}$ of the NV center has been randomized.
} Inspired by this idea, we introduce another interrogation
step after a delay (Steps. 4 and 5 in Fig. \ref{Figure1} b) to extract
the information on how much the environment has been perturbed by
the sensor in the previous steps. These steps include: a rotation
by $\pi/2$ around $y$ direction, interaction for a duration of $t_{\mathrm{I}}$,
and measurement of the $y$ component of sensor spin. Although from
the semi-classical theory, these steps will not produce any signal
as the length of the Bloch vector is $0$ starting from Step. 2, we can
still obtain a signal predicted by the quantum theory. This is the reason \textcolor{black}{why}
this signal is called the \textit{QC} signal.

Now let us investigate the \textit{QC} signal in detail from quantum theory.
In the interrogation process (Step. 1 and Step. 5), the sensor is coupled
to a quantum bath through the Hamiltonian
\[
\hat{V}(t)=\hat{S}_{z}\hat{B}(t),
\]
while decoupled from the bath in other steps. $\hat{B}(t)=e^{i\hat{H}_{\mathrm{B}}t}\hat{B}e^{-i\hat{H}_{\mathrm{B}}t}$
is the time-dependent noise-operator and $\hat{H}_{\mathrm{B}}$ is the
Hamiltonian of the bath. The total system begins with an initial state
$\hat{\rho}_{\mathrm{I}}=\vert x\rangle\langle x\vert\otimes\hat{\rho}_{\mathrm{B}}$.
As shown in Fig. \ref{Figure1} b, the \textit{QC} signal $S_{\mathrm{Q}}$
is the expectation value of the sensor's $\hat{\sigma}_{y}$ in the last
step. The entanglement between the sensor and bath leads to the result (see
supplementary information for details): 
\begin{equation}
S_{\mathrm{Q}}\approx-i\left\langle \left[\hat{\varphi}_{2},\hat{\varphi}_{1}\right]_{-}\right\rangle /2,\label{eq:quantum_signal}
\end{equation}
Here $\left[\right]_{-}$ denotes the commutator and $\left\langle \hat{O}\right\rangle \equiv\mathrm{Tr}_{\mathrm{B}}\left\{ \hat{O}\hat{\rho}_{\mathrm{B}}\right\} /2$.
From this formula, $S_{\mathrm{Q}}$ can thus be attributed to the
commutator of two-phase operators $\hat{\varphi}_{2(1)}\equiv t_{\mathrm{I}}\hat{B}(t_{2(1)})$.
Since classical phases commute to each other, the semi-classical theory
always leads to vanishing \textit{QC} signal $S_{\mathrm{Q}}$. In
contrast to semi-classical theory, quantum theory gives a non-vanishing signal which is proportional to the commutator of the phase operator. 

Eq. ($\ref{eq:quantum_signal}$) forms the basis to remove arbitrary
classical noise since it exists for a quantum entity while vanishes
for any classical noise with an arbitrary pattern. If a classical noise
background $b(t)$ is introduced, the noise operator $\hat{B}(t)$
is changed to $\hat{B}(t)+b(t)$. However, the commutative structure
of $S_{\mathrm{Q}}$ {[}Eq. ($\ref{eq:quantum_signal}$){]} make the
correlation of $b(t)$ absence in \textit{QC} signal $S_{\mathrm{Q}}$\cite{WangCPL2021}.
As a result, the \textit{QC} protocol provides a \textcolor{black}{pattern-independent}
classical noise-free detection of quantum signals.

In comparison with the $\textit{QC}$ protocol, the major difference
between the $\textit{Classical Correlation (CC) }$protocol\cite{LaraouiNC2013}
is that a rotation is added in Step. 2 before the phase randomization process (Fig.
$\ref{Figure1}$ c). This step dramatically changes the physics behind \textcolor{black}{it}.
For$\textit{ QC}$ protocol, the phase acquired in Step. 1, which
contains the environment information, has been eliminated by the phase randomization
process in Step. 2 (Fig. $\ref{Figure1}$ b). As a result, the
$\textit{QC}$ protocol generates the quantum correlation signal as
shown in Eq. ($\ref{eq:quantum_signal}$). As a comparison, for $\textit{ CC}$
protocol, as shown in Fig. $\ref{Figure1}$ c, the added rotation
before the phase randomization process transfers the phase acquired in Step.
1 to the electron population to avoid \textcolor{black}{being} eliminated by the phase randomization
process. Then this phase is correlated to the phase acquired in
Step. 5 to generate the $\textit{CC }$signal. Consequently, the semi-classical
picture still holds in $\textit{ CC}$ protocol. The validity of the semi-classical picture is also indicated in the quantum formula of the $\textit{ CC }$
signal(see supplementary information for details):
\begin{equation}
S_{\mathrm{C}}\approx\left\langle \left[\hat{\varphi}_{2},\hat{\varphi}_{1}\right]_{+}\right\rangle /2.\label{eq:classical_signal}
\end{equation}
which is related to the anti-commutator of two-phase operators. As
a result, if these two-phase operators are replaced by their classical
correspondence $\varphi_{2(1)}$, the\textit{ CC }signal will recover
the semi-classical results $\langle\varphi_{2}\varphi_{1}\rangle$
as shown in Fig. $\ref{Figure1}$ c. This is the reason why it is called the \textit{CC}
signal. The anti-commutative structure of $S_{\mathrm{C}}$ indicates
that the \textit{CC} signal can't filter out the classical background
$b(t)$\cite{WangCPL2021}. Eq. ($\ref{eq:classical_signal}$) also
provides a quantum origin for the classical phase picture which holds
in various nano-scale NMR experiments\cite{LaraouiNC2013,SimonScience2017,BossScience2017,PfenderNC2019,DegenNature2019}.

We illustrate the noise-free detection in the context of nano-scale
magnetic resonance. Here we use the \textit{QC} protocol to extract the
\textit{QC} signal of a single nuclear spin while simultaneously filtering
out arbitrary classical noise. \textcolor{black}{The effective Hamiltonian of the nuclear spin takes the form of $\hat{H}_{\mathrm{B}}=(\omega_{0}+A_{\parallel}/2)\hat{I}_{z}$} and
the coupling to the sensor is $\hat{V}=S_{z}\left[\hat{B}+b(t)\right]$,
where $\hat{B}=A_{\perp}\hat{I}_{x}$ and $b(t)$ is arbitrary classical
noise. The nuclear spin's initial state is set to be $\hat{\rho}_{\mathrm{B}}=1/2+p_{z}\hat{I}_{z}$
and its polarization is $\vert p_{z}\vert\le1$. Eq.($\ref{eq:quantum_signal}$)
leads directly to the \textit{QC} signal for short $t_{\mathrm{I}}$:
\begin{equation}
S_{\mathrm{Q}}\approx\frac{A_{\perp}^{2}t_{\mathrm{I}}^{2}}{4}p_{z}\sin\omega(t_{2}-t_{1})\label{eq:corr_signal_short}
\end{equation}
 where the correlation of classical noise $b(t)$ is absent. However,
the \textit{CC }signal $S_{\mathrm{C}}$ still contains the background
induced by classical noise $b(t)$\cite{WangCPL2021}.

\begin{figure}
\includegraphics[width=\columnwidth]{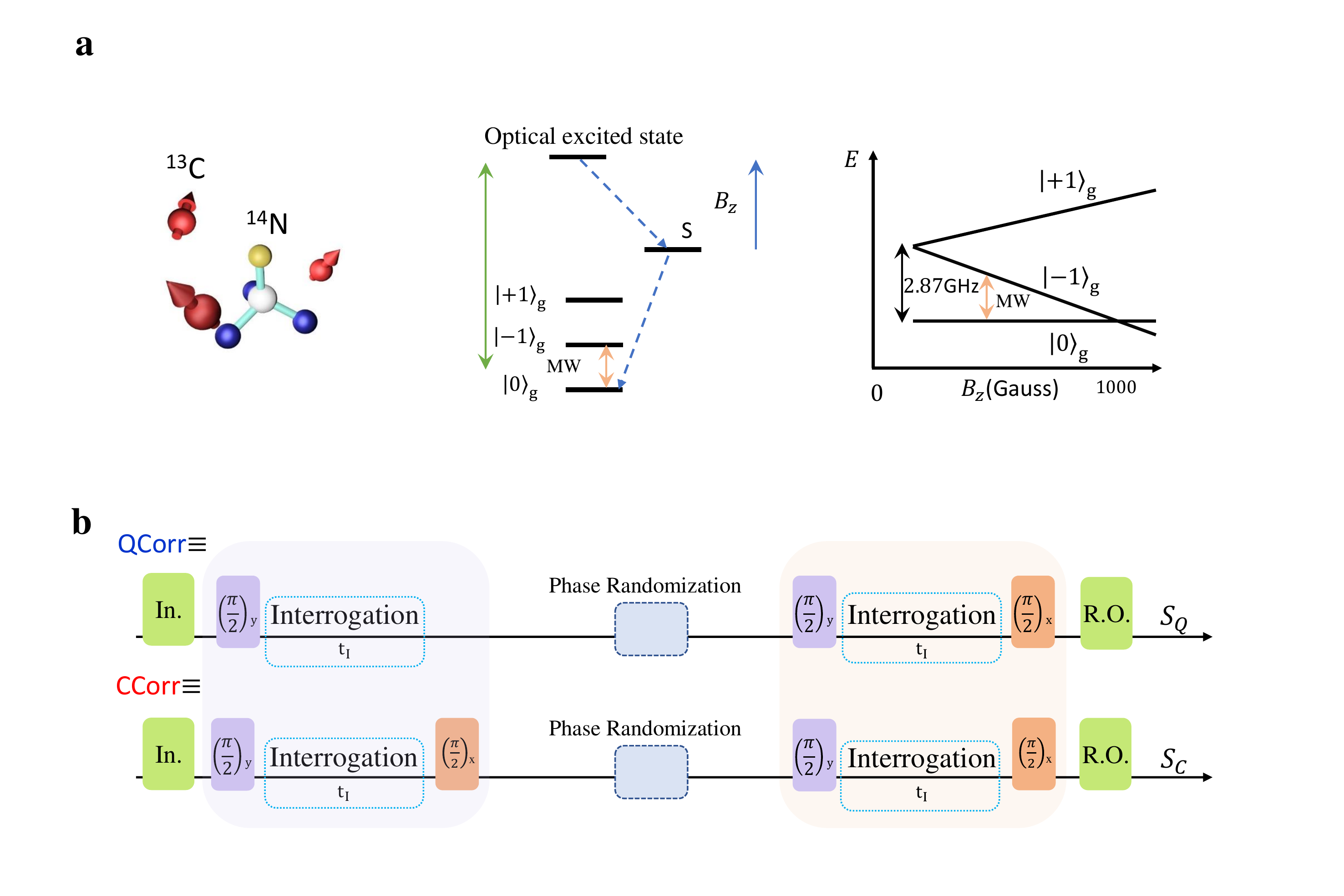}
\caption{\textbf{Experimental implementation of the }\textbf{\textit{QC/CC}}\textbf{
protocol:} $\textbf{a}$ Left graph: The sketch map of the Nitrogen
Vacancy (NV) center in diamond. The NV center is coupled with the surrounding
$^{13}\mathrm{C}$ nuclear spins through dipolar-dipolar interaction;
The middle graph: The optical initialization and operation of the NV center;
The right graph: the energy structure of the ground state of the NV center.
$\textbf{b.}$ The experimental implementation of the protocol. In
these pictures, the green rectangle denotes the laser pulse
to initialize/read out the NV spin. \textcolor{black}{The phase randomization part aims to eliminate the information on the phase of the NV center by randomizing the phase while keeping its population. It is realized by a z-direction rotation DC pulse with random rotating angles (see Supplementary). This method can be generalized to most types of sensors.} The purple(orange) rectangle denotes the $\pi/2$
microwave pulse with axis being $y$($x$).  \label{Figure2}}
\end{figure}

We demonstrate this method in the system of nitrogen-vacancy (NV)
center in diamond with natural $^{13}\mathrm{C}$ abundance and nitrogen
concentration below 3ppb\cite{DohertyNJP2011,JelezkoPSS2006}(see
experimental details in supplementary materials). As shown in the
left graph of Fig. $\ref{Figure2}$ a, the NV center is a negatively charged
deep-level defect in diamond\cite{DohertyNJP2011,JelezkoPSS2006}.
Its ground state is a spin-1 system and has properties such as long
coherence time, high fidelity of initialization, control and readout\cite{DohertyNJP2011}(as
shown in the middle graph of Fig. $\ref{Figure2}$ a). It can coherently
couple to its surrounding individual P1 electron spins and $^{13}\mathrm{C}$
nuclear spins to form quantum computing nodes\cite{ZhaoNatNano2012,KolkowitzPRL2012,TaminiauPRL2012,BradleyPRX2019,TaminiauNature2019};
it can also be used to detect electron spins and nuclear spins in
target molecules outside diamond\cite{DegenNC2021,ShiScience2015}.
Therefore, extraction of the \textit{QC} signal is crucial for its further
development. To demonstrate the proposed protocol, we construct a
quantum sensor by isolating the subspace $\vert0\rangle_{g},\vert-1\rangle_{g}$
of the ground states of the NV center (as shown in the right graph
of Fig. $\ref{Figure2}$ a). Then the protocol defined in Fig. $\ref{Figure1}$
b can be implemented using the pulse sequence shown in Fig. $\ref{Figure2}$
b.

To verify the quantumness of the \textit{QC} signal, we extract\textit{
}it for two different environments surrounding a quantum sensor: 1.
For the classical environment (e.g AC field), the \textit{QC} signal must
vanish due to the absence of intrinsic quantum non-commutability;
2. For the quantum environment, the \textit{QC} signal will exist. 

\begin{figure}
\includegraphics[width=\columnwidth]{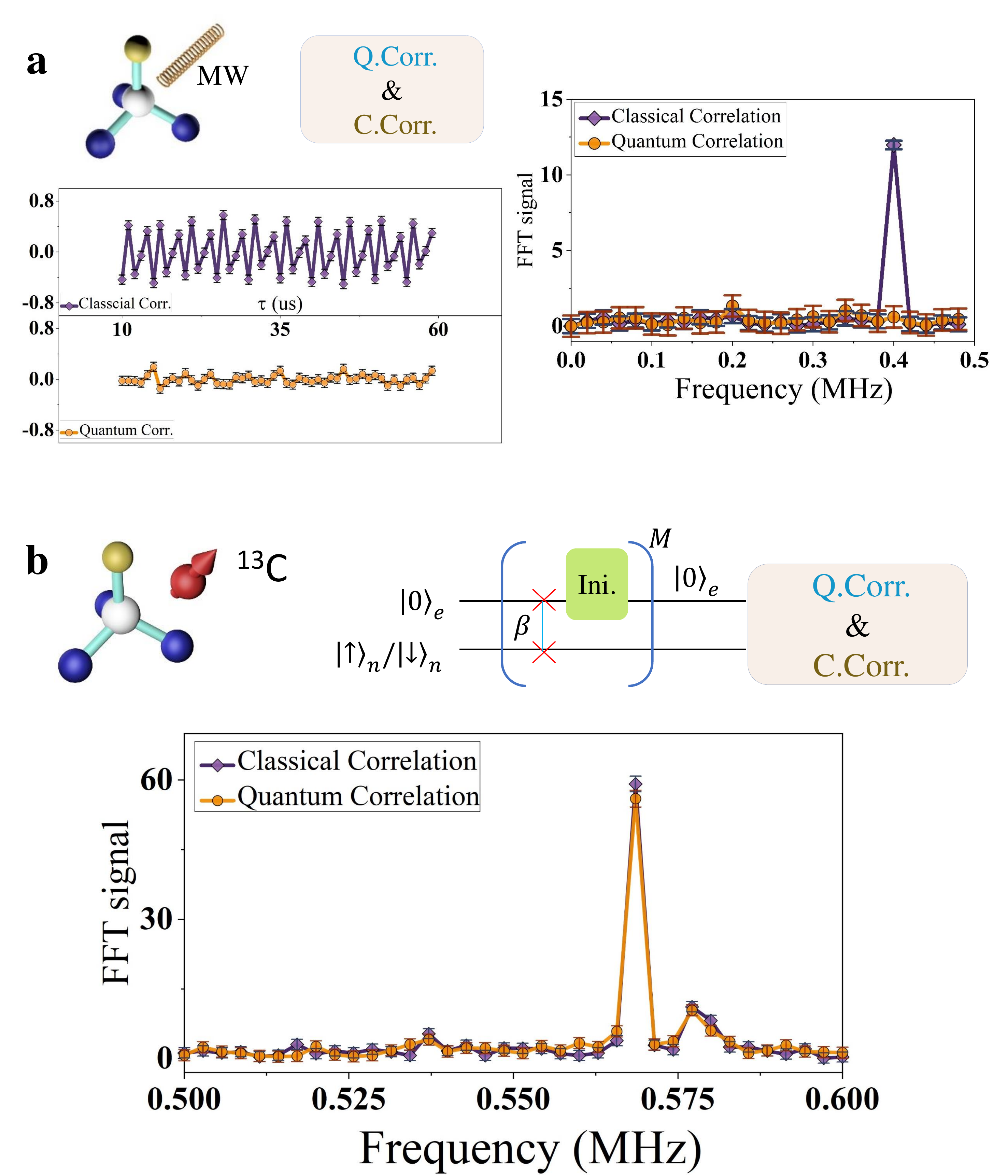}
\caption{\textbf{Verification of the quantumness of }\textbf{\textit{QC }}\textbf{signal.
a.} The \textit{QC/CC} signals from a sensor under the classical AC magnetic
field. Left: the detected \textit{QC} (orange scatters)/\textit{CC} (purple
scatters) correlation signals in the time domain; Right: the detected
\textit{QC/CC} signals in the frequency domain. $\textbf{b.}$ The
\textit{QC/CC} signals from a sensor coupled with nuclear spins. Up:
the sequence to measure \textit{QC/CC} signal of a polarized nuclear
spin. We use the method of Ref. \cite{SchwartzSA2018} to polarize
the nuclear spin bath. The electron polarization is transferred to
the bath by repeating a partial swap gate (denoted by $\beta$) (see
Ref. \cite{SchwartzSA2018}). Down: The Fourier transform of the
\textit{QC/CC} signal. Here the magnetic field is $B_{z}=504\mathrm{G}$. \label{Figure3}}
\end{figure}

For the first aspect, we simulate the classical environment by an
AC magnetic field from a microwave pulse. We use the pulse sequence
in Fig. $\ref{Figure2}$ b to measure both the \textit{QC} and \textit{CC}
signals of an AC magnetic field. The \textit{CC} signal shows a clear
peak while the \textit{QC} vanishes (Fig. $\ref{Figure3}$ a). This
is expected because the classical magnetic field has no intrinsic
non-commutability and hence naturally gives a vanishing \textit{QC}
signal (see Fig. $\ref{Figure1}$ b).

For the second aspect, we detect the\textit{ QC} signal from a polarized
$^{13}\mathrm{C}$ nuclear spin surrounding the NV center spin. Since
the second-order \textit{QC} signal {[}Eq. ($\ref{eq:corr_signal_short}$){]}
vanishes when the bath is in a completely mixed state, we polarize
the $^{13}\mathrm{C}$ nuclear spin by the method introduced in Ref.
\cite{SchwartzSA2018}, where the nuclear spin polarization is transferred
from the polarized electron spin by a series of engineered swap gates (up
graph in Fig. $\ref{Figure3}$ b). In contrast to the absence of the \textit{QC}
signal for the AC field (Fig. $\ref{Figure3}$ a), a clear peak is found
in the Fourier transform of the \textit{QC} signal for the nuclear spin,
which is shown in Fig. $\ref{Figure3}$ b.

The absence of the \textit{QC} signal for classical AC signal indicates
that arbitrary classical background can be filtered out by the \textit{QC}
protocol. In the following, we show how to detect quantum objects free
of classical noise with an arbitrary pattern. We demonstrate it by detecting
remote $^{13}\mathrm{C}$ nuclear spins when two different artificial
noises are applied to the sensor simultaneously. 

\begin{figure}
\includegraphics[width=\columnwidth]{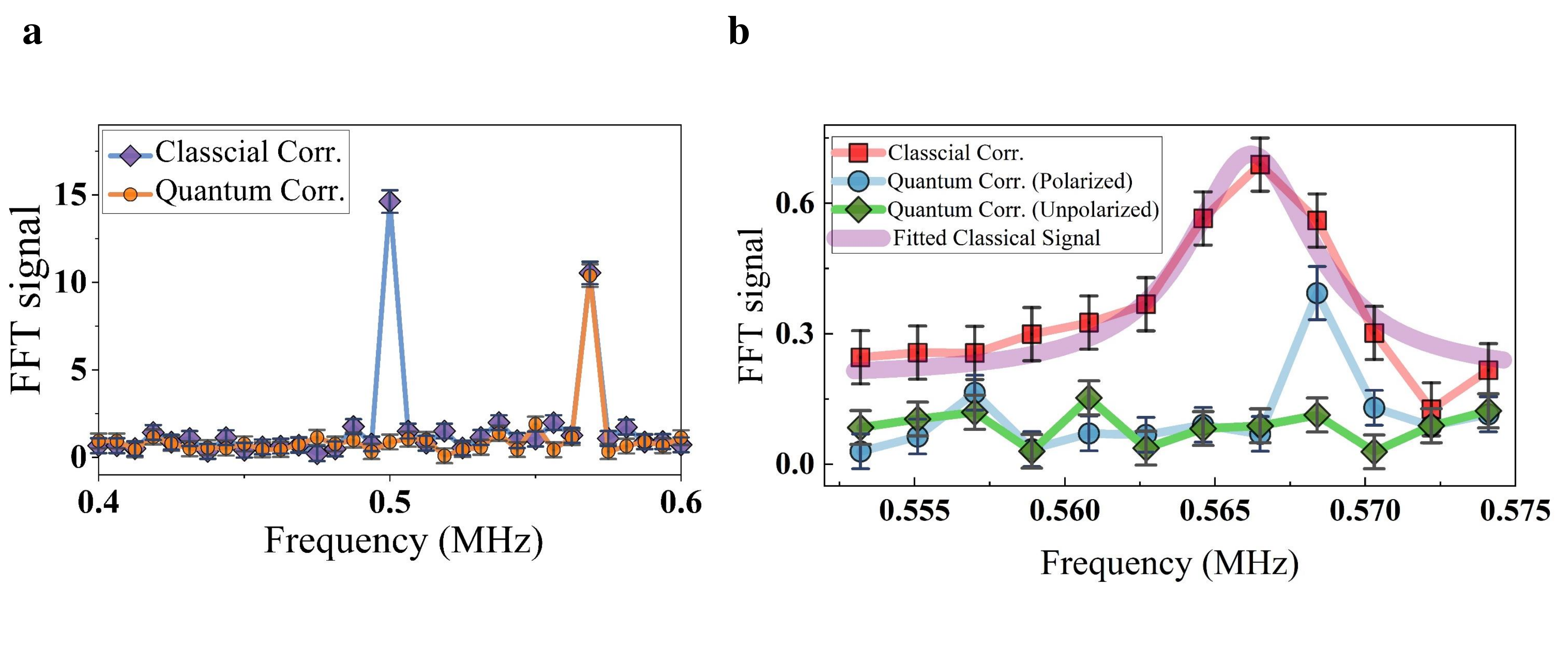}
\caption{\textbf{The demonstration of the pattern-independent quantum sensing:}
$\textbf{a. }$Filter out a single frequency classical noise; $\textbf{b. }$
Filter out classical color noise. Here the magnetic field is $B_{z}=504\mathrm{G}$.
The hyperfine coupling between the nuclear spin and sensor is 60.4
kHz. \label{Figure4}}
\end{figure}

The first case is a narrow bandwidth noise generated by a 500 kHz
AC field at the power of -16dbm. The bandwidth limit\textcolor{blue}{{}
}is 1Hz and the detuning from target nuclear spins is 68.7kHz. As
shown in Fig. $\ref{Figure4}$ a, both the target $^{13}\mathrm{C}$
nuclear spin and its classical noise background occur in the \textit{CC
}signal while only the $^{13}\mathrm{C}$ nuclear spin signal exists
in the quantum one.

The second case is that the classical noise has enough spectral width
(a full-width \textcolor{black}{4.5} kHz at half maximum) and effective spectral strength
to conceal the target signal, under which circumstance the traditional
DSP methods \textcolor{black}{fail}. We generate this artificial color noise with a
Lorentz spectrum shape as indicated by the purple curve in Fig. $\ref{Figure4}$
b (see supplementary information for details). As shown in this figure,
the peak of the \textit{CC }signal is the peak of color noise {[}red
scatters in Fig. $\ref{Figure4}$ b{]} instead of that of the target
nuclear spin. Hence the noise buries the target signal. However, the
\textit{QC} signal (the blue scatters in Fig. $\ref{Figure4}$ b) clearly
singles out the hidden $^{13}\mathrm{C}$ nuclear spins and simultaneously
filters out the classical noise background. 
Although the currently
generated classical noise is a color noise due to the limitation of
the current technique, the quantum nature-based filter method demonstrated
here is in principle suitable for filtering white noise with infinite
spectral width. Therefore this method can be especially useful for
low magnetic field nano-scale NMR\cite{CerrilloPRL2021,PhilipparXiv2021}.

This \textit{QC} protocol works much better when compared with other
DSP methods. Refocusing techniques such as dynamical decoupling have
been widely used in quantum research as bandpass filters. It can
filter out the noise of the first case where signal and noise have
different distributions, but it cannot filter out the noise of the
second case as the classical noise outpowers the quantum one in the
same frequency range\cite{WangCPL2021}. Furthermore, traditional
DSP methods such as active feedback can partially restore the linewidth,
but it is usually hard to recover completely the narrow linewidth
as in the case of the new method demonstrate here. Note that the linewidth
of the nuclear spin signal in Fig. $\ref{Figure4}$ a and the one after
the filtering in Fig. $\ref{Figure4}$ b are similar. This reflects
that the\textit{ QC} protocol can remove the influence of stochastic
classical noise completely. This complete removal and the full restoration
of the linewidth are important for quantum sensing as the linewidth
sets the bound of sensitivity.

The \textit{QC} protocol demonstrated here has the following significance.
First, major DSP methods in quantum science were adapted from the
semi-classical theory, and therefore have not utilized the quantum
non-commuting nature of the target signal to design the filter. As
a result, the method demonstrated here gives the first quantum non-commuting-based DSP method. Second, the detection of the \textit{QC} signal
can also give an unambiguous identification of the quantum environment
and hence gives direct experimental evidence beyond semi-classical
theory. Consequently, these results demonstrate the significance of
\textit{QC} in the open quantum system and also its potential application
in quantum control and sensing. Third, the \textit{QC} protocol uses
a weak-measurement-free approach to single out the quantum signal
which should be absent under the semi-classical theory. Consequently,
it provides a platform-universal prototype for complete characterization
of the quantum environment since the multi-times weak-measurement method
requires a readout technique with high fidelity and high speed\cite{WangCPL2021,WangPRL2019}.
In other words, the technique demonstrated and its potential generalization
are easier to implement in a broad physical qubit system, including
trapped ion/atom, quantum dots, superconducting circuits and defect-based systems, which is important for quantum non-linear sensing\cite{DorfmanRMP2016}.

P.W. is supported by the Talents Introduction Foundation of Beijing Normal University with Grant No.310432106. P.W. \& R.B.L. were supported by Hong Kong RGC General Research Fund (143000119) and NSFC/RGC Joint Research Scheme (N\_CUHK403/16). J.W. acknowledges financial supports from ERC grant SMeL, EU Project ASTERIQS, DFG (GRK2642) and DFG Research group FOR 2724. S.Y. acknowledges financial supports from Hong Kong RGC (GRF/24304617, 14304618).

\bibliographystyle{apsrev4-1}

\end{document}